\documentclass[showpacs,prb,endfloats,preprint]{revtex4}  

\usepackage{graphicx,graphics}
\usepackage{supertabular}%
\usepackage{bm}%
\usepackage{amsmath}%

\begin{document}

\title{Coriolis-Zeeman effect in rotating photonic crystal}

\author{D.L. Boiko}
\email{dmitri.boiko@epfl.ch} \affiliation{\'Ecole Polytechnique
F\'ed\'erale de Lausanne, Quantum Architecture group, 1015,
Lausanne, Switzerland}
\date{ \today }

\begin{abstract}
Rotation-induced splitting of the otherwise  degenerate photonic
bands is predicted for a two-dimensional photonic crystal made of
evanescently coupled microcavities. The symmetry-broken energy
splitting is similar to the Zeeman splitting of atomic levels or
electron's (hole's) magnetic moment sublevels in an external
magnetic field. The orbital motion of photons in periodic
photonic lattice of microcavities is shown to enhance
significantly such Coriolis-Zeeman splitting as compared to a
solitary microcavity [D.L. Boiko, Optics Express \textbf{2}, 397
(1998)]. The equation of motion suggests that nonstationary
rotation induces quantum transitions between photonic states and,
furthermore, that such transitions will serve as a source of
nonstationary gravitational field.
\end{abstract}

\pacs{ 42.70.Qs, 
71.70.Ej, 
03.65.Pm, 
03.30.+p.
}

\maketitle

The Sagnac effect in a rotating ring cavity, known also as the
Coriolis-Zeeman effect for photons, emerges as a frequency
splitting of counterpropagating waves \cite{Heer64}. Thus, for a
ring cavity of $M$ wavelengths optical path, the modal shift  is
$M \Omega$. Nowadays, the effect is used in commercial He-Ne ring
laser gyros of large ($M{\sim} 10^6$) cavity size. Significant
efforts were made towards designing miniature-sized solid-state
devices based on a high optical gain medium \cite{Boiko97}.
Theoretical investigation has been carried out about the impact
of rotation on the whispering gallery modes of a microdisc
microstructure \cite{Nojima0405}. Here, the frequency splitting
scales with the closed optical path length, while the field
polarization is either not relevant or assumed to be parallel to
the rotation axis $\mathbf{\Omega}$.

On the other hand, the Coriolis-Zeeman effect in a cylindrical
microwave resonator rotating along the symmetry axis
\cite{Belonogov69} or in an optical Fabry-P\'erot cavity rotating
in the mirror plane is, at first sight, independent of the cavity
size \cite{Boiko98}. The frequency shift $(S{+}M)\Omega$ of
polarization and transversal modes is set by the spin (${\pm}1$)
and the azimuth mode index $M$. However, the higher is the mode
index $M$, the larger is the size of the cavity that will support
such mode. By virtue of the complexity of the mode discrimination
at high $M$ and because of the small frequency splitting of the
polarization modes, this effect has not yet been verified in
experimental measurements.

\begin {figure}[tbp]
\includegraphics {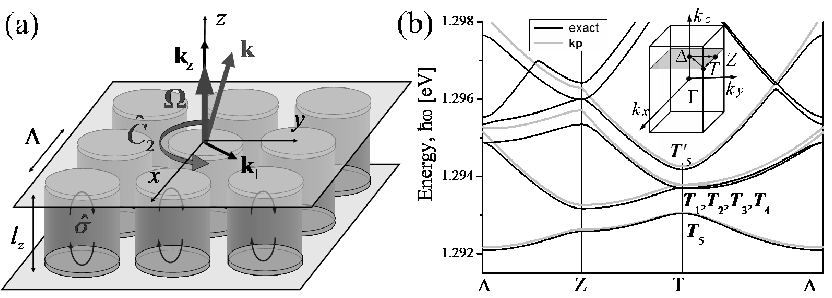}
\caption {(a): Schematic of the array of
coupled microcavities. 
(b): Band structure of square-lattice PhC 
calculated from (\ref{H}) [black curves] and using 8x8
$\mathbf{k{\cdot}p}$  approximation (\ref{4x4UpHkp}) [gray
curves]. The inset shows the BZ and location of the high-symmetry
triangle $\Delta Z T $. The parameters are $\lambda{=}960 nm $,
$n{=}3.53$, $\Lambda{=}4\mu m$, $FF{=}0.65$ and $\Delta \varphi
{=} 0.02$.}\label {figPC_VCSEL} \label {figBandStructure}
\newpage
\end {figure}

In this Letter, I consider the 
Coriolis-Zeeman effect in coupled microcavities arranged in a
periodic two-dimensional (2D) lattice in the plane of rotation
[Fig.\ref{figPC_VCSEL}(a)]. On example of a square-symmetry
lattice, the possibility of enhanced Coriolis-Zeeman splitting,
corresponding to $M {\sim} 1000$, is predicted for the
\emph{low-order} photonic modes \cite{Boiko06}. It is caused by
the photon's orbital motion extended over the large number of
lattice cells. The equation of motion, which is similar to the
Hamiltonian for electrons and holes in magnetic field, suggests
that nonstationary rotation induces quantum transitions between
photonic states  and, vice verse,  that such transitions will
generate a nonstationary gravitational field.

Arrays of evanescently coupled microcavities belong to a
particular sub-class of 2D photonic crystal (PhC) structures
encompassing photonic crystal fibers and arrays of microcavities.
Matrices of vertical cavity surface emitting lasers (VCSELs) are
an example of such 2D photonic crystals \cite{Orenstein91}. In
such structures, only a small transversal component $\mathbf
{k}_\bot$ of the propagation vector $\mathbf{k}$  undergoes Bragg
reflections in the plane of periodic lattice
[Fig.\ref{figPC_VCSEL}(a)]. The structures employ lattices of
periods significantly exceeding the optical wavelength. They are
typically realized by the mirror reflectivity patterning in a
broad-area microcavity. As a model system for such PhC
structures, a Fabry-P\'erot cavity with patterned mirror
reflectivity is considered here. The cavity length is one
wavelength, $l_z{=}\lambda/n $ with $n$ being the refractive
index in the cavity. The reflectivity $R_1(x,y)$ of the one
cavity mirror (\textit{e.g.}, of the top mirror) is modulated in
two directions parallel to its plan. The reflectivity pattern
$R_1{=}\exp i\varphi(x,y)$ consists of pixels with the relative
phase shift $\varphi(x,y){=}\Delta \varphi$ separated by a grid
of $\varphi(x,y){=}0$. The mirror is thus perfectly reflecting
($\left | R_1 \right |{=}1$) and the phase modulation pattern
$\varphi(x,y)$ defines the structure of the cavity modes. The
period of the reflectivity pattern $\Lambda$ is of a few micron
pitch ($\Lambda {\gg} l_z$).
The pattern is characterized by a fill factor $FF$ that is the
ratio between the area of the 
pixel and that of the unit cell. Like in typical VCSEL arrays,
the phase contrast of reflectivity pattern is small ($\left|\Delta
\varphi\right|{\alt} 10^{-2}$). The second mirror of uniform
reflectivity ($R_2{=}1$) has no impact on the cavity modes.

The analysis is carried out here using an equivalent, unfolded
cavity representation \cite{Boiko04}. Multiple reflections at the
cavity mirrors effectively translate the cavity into a structure
that is periodic along the cavity axis (the $z$-axis). The
unfolded PhC is thus three-dimensional and it can be analyzed in
terms of a modal expansion on orthogonal plane waves (OPWs).
The Coriolis-Zeeman effect in such photonic crystal is considered
here using a frame of reference, which rotates together with the
crystal. Such noninertial 
rest frame is
characterized by a metric tensor $g_{ik}$, the off-diagonal
space-time components $g_{0\alpha}$ of which are dependent on the
angular rotation speed $\mathbf{\Omega}$ \cite{LandauII}.
However, in the
unfolded-cavity representation, 
$g_{ik}$ differs from the diagonal Minkowski tensor, even in the
absence of rotation. Thus $g_{0\alpha}{\neq} 0$ at the subsequent
mirror reflections since the reflection operator
$\hat{\sigma}{=}\hat{I}\hat{C}_2$ includes rotation by $\pi$
(about the $z$ axis) followed by coordinate
inversion 
[Fig.\ref{figPC_VCSEL}(a)]. Due to the patterned
 mirror reflectivity,
the equivalent unfolded PhC 
is of periodically varying "noninertiality" in the $xy$ plane
\cite{Boiko04}. In the approximation of the first-order terms
$\Omega r / c$ and $\Delta \varphi$, the metric tensor $g_{ik}$
has  the following nonzero components:
\begin{eqnarray}
g_{00}&{=}&{-}g_{11}{=}{-}g_{22}{=}{-}g_{33}{=}1 \label{metric},  \\
g_{0\alpha}&{=}&{-}\frac1c e_{\alpha\beta\gamma} \Omega^\beta
x^\gamma {-} \delta_{\alpha3}\frac{c}{\omega}
\varphi(x^1,x^2)\sum_{j}\delta (x^3-2jl_z), \nonumber
\end{eqnarray}
where $g_{0\alpha}{=}g_{\alpha0}$ ($\alpha=1,2,3$), the
space-time coordinates are indexed according to the intervals
$dx^0 {=} cdt$, $dx^1 {=} dx$, $dx^2 {=} dy$ and $dx^3 {=} dz$;
twice repeated Greek indexes indicate
summation. 
The first term in $g_{0\alpha}$ accounts for rotation of the
coordinate system \cite{Heer64}. The second term accounts for the
multiple cavity roundtrips along the $z$-axis and reflections at
the cavity mirrors. The $z$-period of the 
unfolded crystal is thus $2l_z$. The metric tensor (\ref{metric})
is validated by inspecting the system Hamiltonian [Eq.(\ref{H})]
for the case of $\varphi{=}0$ (rotating FP cavity \cite{Boiko98})
or $\Omega{=}0$ (PhC in an inertia frame \cite{Boiko04}).

In the approximation (\ref{metric}), the coordinate
space is Euclidean, with the 
metric tensor
$\gamma_{\alpha\beta}{=}-g_{\alpha\beta}+g_{0\alpha}g_{0\beta}/g_{00}
$ being the Kronecker delta $\delta_{\alpha\beta}$. Proceeding in
a standard manner \cite{LandauII}, the covariant Maxwell's
equations with metric (\ref{metric}) are converted to the usual
form in terms of noncovariant field vectors $\mathbf {B}$,
$\mathbf {H}$, $\mathbf {D}$ and $\mathbf {E}$ that assume the
constitutive equations
\begin{equation}
\begin{split}
\mathbf{D} &{=}\varepsilon \mathbf{E}+ \mathbf{H}\times
\mathbf{g}, \quad
\mathbf{B} {=}\mu \mathbf{H} + \mathbf{g}\times \mathbf{E}, \\
\mathbf{g} &{=} \frac{\mathbf{\Omega} \times \mathbf{r}}{c}
+\mathbf {\hat{z}} \frac{c}{\omega }
\varphi(\mathbf{r_{\perp}})\sum_{j}\delta (z-2jl_z),
\end{split}\label{MatEqs}
\end{equation}
where $\mathbf{\hat{z}}$ is the unit vector along $z$-axis
direction and the components of the vector $\mathbf{g}$ are
$g_\alpha{=} -g_{0\alpha}/g_{00}$.

Maxwell' equations in photonic crystal (\ref{MatEqs}) are solved
here by separating fast oscillations in the $z$-axis direction
and slow lateral field oscillations in the $xy$ plane:
\begin{equation}
\left[
\begin{matrix}
E_\alpha \\
H_\gamma
\end{matrix}
  \right]
  {=}e^{ik_z z-i\omega t} \frac{1+\eta(z)}{\sqrt{2\pi}}
\left[
\begin{matrix}
Z^{\frac12} \hat{\mathcal{E}}_{\alpha\beta}\\
Z^{-\frac12}e_{3\beta\alpha}\hat{\mathcal{E}}_{\gamma\alpha}
\end{matrix}
\right] \bm{\psi}_{\beta}(\mathbf{r}_\bot), 
\label{Gauge}
\end{equation}
where $n{=}\sqrt{\varepsilon\mu}$ and $Z{=}\sqrt{\mu /\varepsilon
}$ are the refractive index and impedance in the cavity. The
gauge transformation is introduced here through the operator
\vspace{0.0in}
\begin{equation}
\begin{split}
\hat{\mathcal{E}}_{\alpha\beta}&{=}\delta_{\alpha\beta}\left(1-\frac{1}{4k^2_z}\frac{\partial^2}{\partial
x_\gamma
\partial x_\gamma}
\right) + \frac{1}{2k^2_z}\frac{\partial^2}{\partial x_\alpha
\partial x_\beta} \\
& +i\frac{\delta_{\alpha 3}}{k_z}\frac{\partial}{\partial
x_{\beta} } +\frac \Omega {nc}e_{3\gamma \beta }\left( \delta
_{\alpha 3}x_\gamma -i\frac {x_\alpha}{k_z} \frac \partial
{\partial x_\gamma }\right),
\end{split}\label{OperatorE}
\end{equation}
where the terms ${\sim} k_\bot^2 / k_z^2$ are taken into account.
Such separation of variables is valid in conditions of the
paraxial approximation ($\frac1{k^2_z\left| \bm{\psi}
\right|}\left|\frac{\partial^2 \psi_{\alpha}}{\partial x_{\beta}
\partial x_{\gamma}}\right|
 {\ll} \frac1{k_z\left| \bm{\psi}
\right|}\left|\frac{\partial \psi_{\alpha}}{\partial
x_{\beta}}\right| {\ll} 1$) and of the low contrast of
reflectivity pattern ($\left| \Delta \varphi \right| {\ll} 1$).
The two-component vector $\bm{\psi}(\mathbf{r}_{\bot}){=}{\left(
\begin{smallmatrix} \psi_x \\
\psi_y \end{smallmatrix} \right)}$ in the $x$-$y$ plane is a
slowly-varying function of coordinates. It defines the spatial
patterns of the six electromagnetic field components
(\ref{Gauge}) and it is considered here as the photonic state
wave function. Its squared modulus $\left |
\bm{\psi}(\mathbf{r}_{\bot})\right |^2$ yields the intensity
pattern of the main polarization component in (\ref{Gauge}). For
$\Omega{=}0$, Eqs.(\ref{Gauge})-(\ref{OperatorE}) are in
agreement with the results obtained for the Gaussian
beam \cite{Erikson94}.

In photonic crystals, the wave function
$\bm{\psi}(\mathbf{r}_{\bot})$ is a Bloch wave propagating in the
$xy$ plane \cite{Boiko04},
\begin{eqnarray}
\bm{\psi}_{q\mathbf{k}_\bot} &=& e^{i\mathbf{k}_\bot
\mathbf{r}_\bot}\mathbf{u}_{q\mathbf{k}_\bot}(\mathbf{r}_\bot),
\label{wavefunc}
\end{eqnarray}
where $\frac{4\pi^2}{\Lambda^2} \int_{cell}
\mathbf{u}_{q^{\prime} \mathbf{k}_\bot}^* \mathbf{u}_{q
\mathbf{k}_\bot} d^2\mathbf{r}_\bot {=}
\delta_{q^{\prime}q}$ \cite{Luttinger56}. 
The longitudinal part in (\ref{Gauge}) [the term $e^{ik_z
z}\frac{1+\eta_{q\mathbf{k}}(z)}{\sqrt{2\pi}}$] is also a Bloch
function. Within the $z$-period of the lattice, it has a small
modulation depth $ \left\langle \left|\eta _{q\mathbf{k}}
\right|\right\rangle _{2l_z} {=} \frac{1}{2l_z}\int_{-l_z}^{l_z}
\left|\eta _{q\mathbf{k}} \right | dz {\sim}\Delta \varphi$,
which is set by an effective phase shift $\alpha_{q\mathbf{k}}$
at each reflection of the patterned mirror. The general form of
such periodic function $\eta _{q\mathbf{k}}$ is
\begin{equation}
1{+}\eta _{q\mathbf{k}} {=} \exp
\left\{ i\alpha _{q\mathbf{k}}\sum_{j}\left[ \theta (z{-}2jl_z){-}\frac{1}{2}\right] {-}\frac{iz\alpha _{q\mathbf{k}}}{2l_z}%
\right\}  \label{fmk(z)_aprox}
\end{equation}%
where $\theta(z){=}\int_{-\infty}^z \delta (\zeta)d\zeta $ is the
unit step function. Note that $\eta_{q\mathbf{k}}(z)$ is the odd
function and
$\left\langle \partial \eta _{q\mathbf{k}}/ \partial z%
\right\rangle _{2l_z}{\simeq} 0$ by virtue of the small contrast
of the reflectivity pattern.

By operating with
$e_{3\beta\alpha}\hat{\mathcal{E}}^{-1}_{\gamma\alpha}$ and
$\hat{\mathcal{E}}^{-1}_{\alpha\beta}$ [from (\ref{OperatorE})] on
Maxwell' equations for the curl of $\mathbf{E}$ and
$\mathbf{H}$,  substituting the gauge (\ref{Gauge}) and averaging
over the $z$-period of the lattice, the Maxwell' equations are
converted into the same form of a Hamiltonian eigenproblem with
respect to the photonic state wave function
$\bm{\psi}_{q\mathbf{k}}(\mathbf{r}_{\bot })$
\begin{equation}
\begin{split}
&\left( \frac{m_{0} c^{2}}{n^{2}}+\frac{\mathbf{\hat{p}}_{\bot
}^{2}}{2m_{0}} - \frac{c\hbar }{2nl_z}\varphi(\mathbf{r}_{\bot
})\right)
\bm{\psi}_{q\mathbf{k}} \\
& \qquad - \frac{\Omega}{n^2}\left(\mathbf{r}_{\bot }\times
\mathbf{\hat{p}}_{\bot } + \hbar\hat{S}_z \right)
\bm{\psi}_{q\mathbf{k}} =  \hbar \omega _{q\mathbf{k}}
\bm{\psi}_{q\mathbf{k}},
\end{split}\label{H}
\end{equation}%
where $m_0{=}n\hbar k_z/c $ and
$\hat{S}_z{=}i\mathbf{\hat{z}}\times$ is the spin operator that
reads $(\hat S_z)_{\alpha\beta}{=}i e_{\alpha 3\beta}{=}\left(
\begin{smallmatrix} 0 & -i \\
i & 0
\end{smallmatrix}
\right)$ in the basis of the two-component vector functions
$\bm{\psi}{=}\left(
\begin{smallmatrix} \psi_x \\
\psi_y \end{smallmatrix}
\right)$.
Solutions of Eq.(\ref{H}) define the slowly-varying 
components of photonic modes (\ref{Gauge}). The difference
between the exact equation for
$(1{+}\eta_{q\mathbf{k}})\bm{\psi}_{q\mathbf{k}}$ and its
$z$-period average [Eq.(\ref{H})] yields the equation for the
periodic part of the fast longitudinal component:
\vspace{0.00in}
\begin{equation}
\frac{\partial \eta _{q\mathbf{k}}}{\partial z} = i\alpha
_{q\mathbf{k}}\left( 1+\eta _{q\mathbf{k}}\right) \left[
\sum_{j}\delta (z-2jl_z)-\frac{1}{2l_z}\right] \label{eta}
\end{equation}
where $\alpha_{q \mathbf{k}} {=} \left\langle \mathbf{\psi}
_{q\mathbf{k}} \right| \varphi \left|\mathbf{\psi} _{q\mathbf{k}}
\right \rangle$.
The solution of (\ref{eta}) is given by (\ref{fmk(z)_aprox})
provided that the eigen functions $\bm{\psi}_{q\mathbf{k}}$ of
the Hamiltonian (\ref{H}) are known.

In (\ref{H}), the first term is due to the paraxial propagation
along the $z$ axis. The second and third terms are the in-plane
kinetic energy and the periodic crystal potential, respectively.
The last term in (\ref{H}) is a perturbation induced by the
Coriolis force. For $\varphi{=}0$ and $\Omega{=}0$, within the
accuracy of the time variable, Eq.(\ref{H}) is just the paraxial
wave equation. For $\varphi{=}0$, Eq.(\ref{H}) yields the
effective refractive index $ \frac{c}{\omega}(k_z {+} \frac
{k_{\bot}^2}{2k_z})$ of a circularly-polarized paraxial wave
\begin{equation}
n_{\text{eff}}=n+\bm{g}\bm{\tau}\pm\frac{\bm{\Omega}\bm{\tau}}{\omega
n}, \label{ref_index}
\end{equation}
where $\bm{\tau}{=}\frac{\mathbf{k}}{k}$ defines the propagation
direction, ${+}$/${-}$ sign is for the left/right handed
polarization. Eq.(\ref{ref_index}) agrees with previously
reported expressions for the axial nonreciprocity \cite{Heer64}
(second term)
and circular birefringence \cite{Boiko98} (third term) induced by
the Coriolis force for photons. Finally, the unperturbed
($\Omega{=}0$) Hamiltonian has been verified by
experimental measurements in VCSEL array PhC 
heterostructures \cite{Guerrero04}. These justify the
approximation (\ref{H}) of the 
Hamiltonian.

\begingroup
\squeezetable
\begin{table}[tbp]
\caption{Basis functions (scalars) and photonic harmonics
(vectors) of irreducible representations of the group of $\mathbf
k$ at the $T$ point of the BZ ($C_{4v}$ point group)}
\label{PCH_C4v_new}
\begin{ruledtabular}
\begin{tabular}[t]{llll}
$ T_i $ &  & $T_{i}\times T_{5} $ &  \\
\hline
$T_{1}$ & $S$ & $\mathbf{T}_{5}$ & $S \mathbf{\hat{x}}, S \mathbf{\hat{y}}$\\
$T_{2}$ & $XY(X^{2}-Y^{2})$ & $\mathbf{T}_{5}^{\prime\prime}$ & $XY(X^{2}-Y^{2})\mathbf{\hat{x}},XY(X^{2}-Y^{2})\mathbf{\hat{y}}$ \\
$T_{3}$ & $X^{2}-Y^{2}$ & $\mathbf{T}_{5}^{\prime\prime\prime}$ & $(X^{2}-Y^{2})\mathbf{\hat{x}},(X^{2}-Y^{2})\mathbf{\hat{y}}$ \\
$T_{4}$ & $XY$ & $\mathbf{T}_{5}^\prime$ & $XY\mathbf{\hat{x}},  XY\mathbf{\hat{y}}$ \\
$T_{5}$ & $iX,iY$ & \multicolumn{2}{l}{
$\mathbf{T}_{1}+\mathbf{T}_{2}+\mathbf{T}_{3}+\mathbf{T}_{4}:$}
\\
& & $ \mathbf{T}_{1}$ & $\frac{i}{\sqrt2}\left(X\mathbf{\hat{x}}+Y\mathbf{\hat{y}}\right) $\\
& & $\mathbf{T}_{2}$ & $\frac{i}{\sqrt2}\left(Y\mathbf{\hat{x}}-X\mathbf{\hat{y}}\right)$ \\
& & $\mathbf{T}_{3}$ & $\frac{i}{\sqrt2}\left(X\mathbf{\hat{x}}-Y\mathbf{\hat{y}}\right) $\\
& & $\mathbf{T}_{4}$ &
$\frac{i}{\sqrt2}\left(Y\mathbf{\hat{x}}+X\mathbf{\hat{y}}
\right)$
\end{tabular}
\end{ruledtabular}
\end{table}
\endgroup

Analytical similarities between the effective single-electron
Hamiltonian in a semiconductor subjected to an external magnetic
field
and 
Eq.(\ref{H}) allow the correspondence between the periodic crystal
potential $V$ and phase 
pattern 
$\varphi$ ($ V {\rightarrow} {-} \frac{c\hbar }{2nL}\varphi $),
and the vector potentials $\mathbf A{=} \frac 12 \mathbf{H
{\times} r} $ and
$\mathbf{g}_\bot{=}\frac 1c \mathbf{\Omega} {\times} \mathbf{r} $
($ \frac e c \mathbf A {\rightarrow} \frac{m_0c}{n^2} \mathbf
{g}_\bot
 $).
Photons in 
photonic crystal subjected to a nonpermanent gravitatinoal field
exhibit thus a similar behaviour with electrons (holes) in a
magnetic field. Accordingly, the impact of rotation on the
envelope function and periodic part of the photonic Bloch wave
(\ref{wavefunc}) is different. Like in the case of electrons,
\cite{Luttinger56} the components of velocity operator
$\mathbf{\hat v}_\bot{=}\frac 1{i\hbar} [\mathbf{r}_\bot, \hat
H]$ do not commute ($\left[ \hat v_x, \hat
v_y\right]{=}-\frac{2i\hbar}{m_0 n^2} \Omega_z$, where
$\mathbf{\hat v}_\bot{=} \frac {\mathbf{\hat p}_\bot}{m_0} {-}
\frac {\mathbf \Omega \times \mathbf r_\bot}{n^2}  $). However,
the second-order terms ${\sim}\frac {\Omega^2 r^2} {c^2}$ in the
Hamiltonian (\ref{H}) [and, respectively, in
(\ref{metric})] are needed to define whether the Landau-like
quantization is possible for photonic
envelope wave functions. 
In the rest of the Letter, the impact of rotation on the periodic
part of Bloch functions is examined in details on example of a
square-lattice PhC.

\begin {figure}[tbp]
\includegraphics {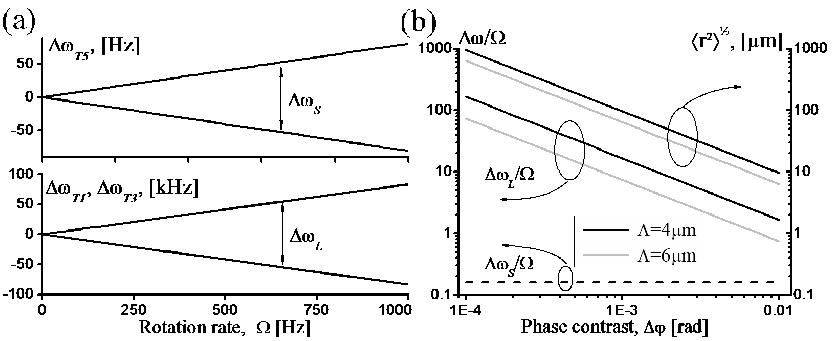} 
\caption {(a): Coriolis-Zeeman splitting of the $\mathbf {T}_5$
($\mathbf {T^\prime}_5$) (top panel) and $\mathbf {T}_1$-$\mathbf
{T}_4$ (bottom panel) bands as a function of rotation rate
$\Omega$. $\Lambda = 4\mu m $ and $\Delta \varphi=10^{-4}$. (b):
Relative splitting $\frac {\Delta \omega_{L,S}} {\Omega} $ (left
axis) and matrix element $\langle \mathbf T_1 | \mathbf {r}^2 |
\mathbf T_1 \rangle ^{\frac 12}$ (right axis) as a function of the
lattice contrast $\Delta \varphi$. The lattice constant $\Lambda$
is $4$ (black curves) and $6\mu m$ (gray curves), $n=3.53$. Other
parameters are given in the caption of Fig.\ref{figBandStructure}.
}\label {figEffects} \label
{figZeemanScale}
\newpage
\end {figure}

Fig.\ref{figBandStructure}(b) shows the typical band structure of
a square-lattice PhC, which is calculated along the high symmetry
lines $\Delta$-$Z$-$T$ in the Brillouin zone (BZ), using the OPW
expansion in unperturbed ($\Omega{=}0$) Hamiltonian. By virtue of
the square lattice symmetry, all states are degenerate by the
 photon's spin (\textit{e.g.},
the doubly degenerate states $\mathbf{T}_5$ or
$\mathbf{T}^\prime_5$). Angular rotation removes the degeneracy
of such states and splits their energies on
$2\frac{\hbar\Omega}{n^2}$ [Fig.\ref{figEffects}(a), top panel].
However, there are states,
like the degenerate states $\mathbf{T}_1$, $\mathbf{T}_2$,
$\mathbf{T}_3$ and $\mathbf{T}_4$, of the four-fold degeneracy,
which is caused by the orbital symmetry of the Bloch functions.
For such states, there is an important angular momentum
contribution $\frac{1}{n^2}\hbar\Omega_z \hat{L}_z $ to
the 
energy shift.

The Coriolis-Zeeman splitting of 
these states is analyzed here using the first-order $\mathbf{kp}$
expansion in the $T$ point [$\mathbf{k}=(\frac \pi \Lambda ,
\frac \pi \Lambda, k_z )$] of the BZ. The expansion basis is
deduced from the empty lattice test. Four scalar plane waves
$\exp({\pm} \frac {\pi x} \Lambda {\pm} \frac {\pi y} \Lambda )$,
which form the first photonic band of empty lattice, originate
from the nearest equivalent $T$ points of reciprocal lattice and
provide representation that is reducible under the $C_{4v}$ point
group (the symmetry group of $\mathbf{k}$). Their symmetrized
combinations of $T_1$, $T_4$ and $T_5$ representations are
indicated in the second column of Table \ref{PCH_C4v_new} with
the capitals letters corresponding to the main term of Taylor
expansion on parameter $\frac{|\mathbf{r}_\bot|} {\Lambda}$
(\textit{e.g.}, $\left|S\right\rangle {=}\frac{1}{\pi} \cos \frac
{\pi x} \Lambda \cos \frac {\pi y} \Lambda  $,
$\left|iX\right\rangle {=}\frac{i}{\pi} \sin \frac {\pi x}
\Lambda \cos \frac {\pi y} \Lambda  $). The photon's spin
transforms as the two-dimensional representation $T_5$.
Therefore, a reduction of the direct product $T_i\otimes
T_5$ 
results in the eight symmetry adapted photonic harmonics of
$\mathbf T_{1}$-$\mathbf T_{5}$ and $\mathbf T^\prime_5$
representations that constitute a suitable $\mathbf{k}
\mathbf{p}$-expansion basis in the low-order photonic bands.
These states are well separated energetically from the other
states in the $T$ point [Fig.1(b)].

For a general state $\left| \bm{\psi}_{q\mathbf{k}} \right
\rangle{=}e^{i \mathbf{k_\bot r}_\bot } \sum_i c_i \left |
\mathbf{T}_i \right \rangle$  at $\mathbf {k}_\bot$ measured from
the $T$ point of the BZ,  the 8x8 $\mathbf{kp}$ Hamiltonian for
the coefficients $c_i$ is of the block-diagonal form
\begin{equation}
\hat{H}=\left[
\begin{smallmatrix}
\hat{H}_0+\hat{H}_{\mathbf{k}\mathbf{p}}+\hat{H}_\Omega
+\frac{\hbar
^2k_{\bot }^2}{2m_0} & 0  \\
0 & \hat{H}_0+\hat{H}_{\mathbf{k} \mathbf{p}}^{*}-\hat{H}_\Omega
^{*}+\frac{\hbar ^2k_{\bot }^2}{2m_0}
\end{smallmatrix}
\right]\label{4x4UpHkp}
\end{equation}
where
\begin{eqnarray}
\!\!\!\!\!\!&&\hat{H}_0{=}\left[
\begin{smallmatrix}
\hbar \omega_{T_5^\prime} & 0 & 0 & 0 \\
0 & \hbar \omega_{T_1} & 0 & 0  \\
0 & 0 & \hbar \omega_{T_1} & 0 \\
0 & 0 & 0 & \hbar \omega_{T_5}
\end{smallmatrix}
\right]\!\!\!,   \hat{H}_{\mathbf{k}\mathbf{p}}{=} \frac{\hbar
P}{m_0} \!\!\left[
\begin{smallmatrix}
0 & k_{-} & k_{+} & 0 \\
k_{+} & 0 & 0 & k_{-}  \\
k_{-} & 0 & 0 & -k_{+} \\
0 & k_{+} & -k_{-} & 0
\end{smallmatrix}
\right]\!\!,
\nonumber \\
\!\!\!\!\!\!&&\hat{H}_\Omega {=} {-} \frac{\hbar \Omega }{n^2}
\left[
\begin{smallmatrix}
1 & -M_{-}\frac {\hbar k_{-}} {2P} & M_{-}\frac {\hbar k_{+}} {2P} & 0 \\
-M_{-}\frac {\hbar k_{+}} {2P} & -M+1 & 0 & -M_{+}\frac {\hbar k_{-}} {2P} \\
M_{-}\frac {\hbar k_{-}} {2P} & 0 & M+1 & -M_{+}\frac {\hbar k_{+}} {2P} \\
0 & -M_{+}\frac {\hbar k_{+}} {2P} & -M_{+}\frac {\hbar k_{-}}
{2P} & 1
\end{smallmatrix}
\right] \nonumber
\end{eqnarray}
in the basis of functions $\begin{smallmatrix}
\frac 1{\sqrt{2}}(\mathbf{T}_{5\mathbf{x}%
}^{\prime }\pm i\mathbf{T}_{5\mathbf{y}}^{\prime }) \end{smallmatrix}$, $\begin{smallmatrix} \frac{ \mp i}{\sqrt{2}}%
\left( \mathbf{T}_1 \mp i\mathbf{T}_2\right) \end{smallmatrix}$, $\begin{smallmatrix} \frac { \pm i}{\sqrt{2}}\left( \mathbf{%
T}_3 \pm i\mathbf{T}_4\right) \end{smallmatrix}$, $\begin{smallmatrix} \frac{\mp i}{\sqrt{2}}(\mathbf{T}_{5\mathbf{x}} \pm i%
\mathbf{T}_{5\mathbf{y}})
 \end{smallmatrix}$ \cite{ExpliciteBasis}.
The upper (lower) sign refers to the top (bottom) 4x4 block. In
(\ref{4x4UpHkp}), $k_{\pm}{=}k_x \pm i k_y$, $P{=}$
$\frac{1}{\sqrt{2}}\langle S |{-}i\hbar \frac{\partial}{\partial
x}| iX \rangle{=}\frac{\hbar \pi }{\sqrt 2 \Lambda } $ is the
interband matrix element of $\mathbf{\hat{p}}$ that
defines the band mixing. The eigen solutions of (\ref{4x4UpHkp})
are plotted in Fig.1(b). Good agreement with the band structure
calculated from Eq.(\ref{H}) justifies the first-order
$\mathbf{kp}$ approximation
(\ref{4x4UpHkp}) of the system Hamiltonian.

The matrix $\hat{H}_\Omega$ in (\ref{4x4UpHkp}) is a perturbation
induced by rotation. It accounts for the Coriolis-Zeeman energy
shift, which is of the opposite sign for the left (upper 4x4
block) and right (lower block) handed polarization states of the
same orbital symmetry.
The rotation lifts degeneracy between such 
spin states, producing 
the frequency splitting $\Delta \omega_S=2\frac{\Omega}{n^2}$, as
indicated in Fig.\ref{figEffects}(a) [top panel] on example of
the $\mathbf {\hat T}_5$ ($\mathbf {\hat T^\prime}_5$) states. In
(\ref{4x4UpHkp}), the parameter $M{=}$ $M_{+} {+} M_{-}$ accounts
for the orbital part of wave functions. The orbital contribution
$\Delta \omega_L{=}2\frac{M\Omega}{n^2}$ to the Coriolis-Zeeman
energy shift ${\pm} \frac 12 \hbar\Delta \omega_L {\pm} \frac 12
\hbar\Delta \omega_S$ in the $\mathbf{T}_{1}$-$\mathbf{T}_{4}$
states is shown in the bottom panel of Fig.\ref{figEffects}(a)
($\Delta \omega_S$ is not visible at the scale of $\Delta
\omega_L$). The energy shift $\pm \frac 12 \hbar\Delta \omega_L$
[the term ${-}\frac{\mathbf{\Omega}}{n^2} \mathbf {r} {\times}
\mathbf {\hat p}$ in Eq.(\ref{H})]  is evaluated here using the
relationship $\mathbf r_{mn}{=}\frac {i \hbar}{m_0} \frac
{\mathbf p_{mn}}{E_n-E_m} $. Note that the procedure to evaluate
the matrix elements of $\hbar \hat L_z{=}
\mathbf{r}_\bot {\times} \mathbf{ \hat p}_\bot$ 
in free space \cite{ONeil02} and in periodic lattices
\cite{Bir_Pikus} is different. Here, the \textit{f}-sum rule
$\frac{1}{m_{\alpha\beta}}{=}\frac{\delta_{\alpha\beta}}{m_0}{+}\frac{2}{m_0^2}\sum_n
\frac{p_{mn}^{\alpha}p_{nm}^{\beta}}{E_n-E_m}$ \cite{Luttinger56}
implies that $M_{\pm}=\mp\frac 12 (
\frac{m_0}{m_{{T}_5,{T}_5^{\prime}}}{-}1)$ and
\vspace{0.0in}
\begin{equation}
\Delta \omega_L=\frac{\mathbf{\hat{z} \Omega}}{n^2} \left[
\frac{m_0}{m_{{T}_5^\prime}}{-}\frac{m_0}{m_{{T}_5}}\right],
\qquad \Delta \omega_S= 2\frac{\mathbf{\hat{z} \Omega}} {n^2}.
\label{DeltaOmega}
\end{equation}
For an array of square pixels defining the microcavities,
$M_{\pm}{=}\frac{2nl_zP^2}{\hbar m_0 c FF\Delta \varphi } \left[
\frac{\sin \pi \sqrt{FF} }{\pi \sqrt{FF}}( 1{\pm}\frac{\sin \pi
\sqrt{FF}}{\pi \sqrt{FF}}) \right]^{-1} $. It follows that
reducing the effective mass (via the lattice pitch $\Lambda$,
fill factor $FF$ and contrast $\Delta\varphi$), one can enhance
the Coriolis-Zeeman splitting and achieve $M$ of more than
$10^3$. [Fig.\ref{figZeemanScale}(b) shows the ratio $\frac{\Delta
\omega_L}{\Omega}=\frac{2M}{n^2}$, left axis]. The enhancement is
caused by the weak localization of photonic wave functions to the
lattice sites.
The intraband 
matrix element $\langle q \mathbf k | \mathbf {r}^2 |  q \mathbf
k \rangle
 {=}\frac
{\hbar^2(M_{-}^2{+}M_{+}^2)}{2 P^2} $ in the $\mathbf
T_1$-$\mathbf T_4$ bands indicates that the 
wave functions (and, hence, the photon's orbital motion) spread
over a large PhC crystal domain [Fig.\ref{figZeemanScale}(b)
right axis], such  that $\frac{ \Delta \omega_{L}}{\Omega }{=}
\frac{2\pi\sqrt{2\langle\mathbf{r}_\bot^2 \rangle
}}{n^2\Lambda}[1{+}\text{sinc}^2\pi\sqrt{FF}]^{-1}$. In the bands
of $\mathbf{T}_5$ symmetry, the intraband matrix elements of
$\mathbf{r}^2$ are nonzero as well, however, the orbital
contribution to the Coriolis-Zeeman energy vanishes, in
accordance with the group theory selection rules. With present
experimental techniques, the predicted frequency splitting
[Fig.\ref{figZeemanScale}(a)] can be validated by direct
measurements.

The analogy between Eq.(\ref{H}) and electron's (hole's) magnetic
moment Hamiltonian suggests that a nonstationary field
$\mathbf{g}(t)$ will stimulate transitions between the photonic
states, and that such transitions will serve as a source for
nonpermanent gravitational field $\mathbf{g}(t)$.

The author is grateful to Eli Kapon, Marc-Andr\'e Dupertuis and
Edoardo Charbon, as well as to the Fonds National Suisse de la
R\'echerche Scientifique.

\newpage


\begin{thebibliography}{15}
\bibitem{ExpliciteBasis} The wave functions are
$\begin{smallmatrix}\frac 1{\sqrt{2}}XY\left( \mathbf{\hat{x}}
\pm i\mathbf{\hat{y}}\right) \end{smallmatrix}$,
$\begin{smallmatrix} \pm \frac 12(X \mp iY)(\mathbf{\hat{x}} \pm i
\mathbf{\hat{y})} \end{smallmatrix}$, $\begin{smallmatrix} \mp
\frac{1}2(X \pm iY)(\mathbf{\hat{x}} \pm i\mathbf{\hat{y})}
\end{smallmatrix}$, $\begin{smallmatrix} \mp \frac{
i}{\sqrt{2}} S \left( \mathbf{\hat{x}} \pm
i\mathbf{\hat{y}}\right)
\end{smallmatrix}$.
\end{thebibliography}
\end{document}